\documentclass[11pt]{article}
\usepackage{graphicx}
\usepackage{cite}

\begin{document}

%
%
\title{\Large\bf 
Transverse spin asymmetries for {\boldmath $W$}-production in \\
proton-proton collisions}

\author{Andreas Metz, Jian Zhou
 \\[0.3cm]
{\normalsize\it Department of Physics, Barton Hall,
  Temple University, Philadelphia, PA 19122, USA}}

\maketitle


%
%
\begin{abstract}
\noindent
We study parity-even and parity-odd polarization observables for the process 
$p \, p \to l^{\pm} \, X$, where the lepton comes from the decay of a $W$-boson.
By using the collinear twist-3 factorization approach, we consider the case 
when one proton is transversely polarized, while the other is either 
unpolarized or longitudinally polarized.
These observables give access to two particular quark-gluon-quark correlation 
functions, which have a direct relation to transverse momentum dependent 
parton distributions.
We present numerical estimates for RHIC kinematics.
Measuring, for instance, the parity-even transverse single spin correlation 
would provide a crucial test of our current understanding of single spin 
asymmetries in the framework of QCD.
\end{abstract}

%
%
\section{Introduction}
\noindent
It has long been recognized that production of $W$-bosons in hadronic 
collisions can provide new insights into the partonic structure of hadrons, 
with polarization observables being of particular interest.
In this context the parity-odd longitudinal single spin asymmetry (SSA) in 
proton-proton scattering plays a very important role, both for leptonic as 
well as hadronic final states 
(see~\cite{Bourrely:1990pz,Bourrely:1993dd,Weber:1993xm,Kamal:1997fg,Gehrmann:1997ez,Bunce:2000uv,Gluck:2000ek,Nadolsky:2003fz,Moretti:2005aa,Arnold:2008zx,Berger:2008jp,deFlorian:2010aa} 
and references therein).
A major aim of looking into this observable is to get new and complementary
information on the quark helicity distributions inside the proton.

In the meantime, also a few studies for $W$-production with transversely 
polarized protons 
are available~\cite{Brodsky:2002pr,Schmidt:2003wi,Kang:2009bp,Kang:2010fu}.
These papers mainly focus on a particular parity-even transverse single 
spin effect in $p \, p \to W^{\pm} \, X$ (with a subsequent decay of the $W^{\pm}$ 
into a lepton pair) that is related to the transverse momentum 
dependent Sivers function $f_{1T}^\perp$~\cite{Sivers:1989cc} in the polarized proton.
Such an observable could, in principle, be measured at the Relativistic 
Heavy Ion Collider (RHIC) in Brookhaven.
In order to have clean access to transverse momentum dependent parton distributions 
(TMDs) like the Sivers function, one has to reconstruct the $W$-boson in the experiment.
However, what one measures is $p \, p \to l^{\pm} \, X$, and the detectors at 
RHIC do not allow to fully determine the momentum of the $W$.

The kinematics for inclusive production of a single lepton in proton-proton 
collisions coincides with the one for inclusive production of a jet or a hadron,
for which mostly collinear factorization is used in the literature.
In this Letter, we compute transverse spin observables for $p \, p \to l^{\pm} \, X$
in the collinear twist-3 formalism at the level of Born diagrams.
The machinery of collinear twist-3 factorization was pioneered already in the 
early 1980's~\cite{Efremov:1981sh,Ellis:1982wd}, and in the meantime frequently 
applied to transverse spin effects in hard semi-inclusive reactions
(see~\cite{Efremov:1984ip,Qiu:1991pp,Eguchi:2006mc,Zhou:2009jm} and references 
therein).

If one of the protons in $p \, p \to l^{\pm} \, X$ is transversely polarized, 
and the other is either unpolarized or longitudinally polarized, one can identify 
two parity-even and two parity-odd spin observables.
We will discuss below that, in the collinear twist-3 approach, these four 
observables contain two specific twist-3 quark-gluon-quark correlation functions.
One is the so-called ETQS (Efremov-Teryaev-Qiu-Sterman) matrix 
element~\cite{Efremov:1981sh,Efremov:1984ip,Qiu:1991pp}, which is related to a
particular moment of the transverse momentum dependent Sivers function as shown 
in~\cite{Boer:2003cm,Ma:2003ut}.
The second is related to the TMD $g_{1T}$~\cite{Zhou:2008mz,Zhou:2009jm}, 
where we use the TMD-notation of 
Refs.~\cite{Mulders:1995dh,Boer:1997nt,Bacchetta:2006tn,Arnold:2008kf}.

In addition to the analytical results, we provide numerical estimates for
typical RHIC kinematics $(\sqrt{s} = 500 \, \textrm{GeV})$.
All the observables are peaked around $l_T \approx M_W/2$, with $l_T$ 
representing the transverse momentum of the lepton and $M_W$ the $W$-boson 
mass.
In each case we predict clearly measurable effects.
For the parity-even transverse SSA $A_{TU}^{e}$ our numerical results are very 
close to those obtained in Ref.~\cite{Kang:2009bp} on the basis of factorization 
in terms of transverse momentum dependent parton correlators. 

Before presenting our results we emphasize that measuring $A_{TU}^{e}$ would 
provide a crucial test of our present understanding of transverse SSAs in QCD.
In particular, this means that such a measurement would test the same physics 
--- the gluon exchange between the remnants of the hadrons and the active 
partons --- which underlies the famous process-dependence of the Sivers function 
and of related time-reversal odd parton distributions~\cite{Collins:2002kn}.
In other words, experimental results for $A_{TU}^{e}$ in $p \, p \to l^{\pm} \, X$,
even if analyzed in terms of collinear parton correlators, would check a crucial 
ingredient of TMD-factorization~\cite{Collins:1981uk,Ji:2004wu,Collins:2004nx}. 
Such a check, in essence, can be considered to be as fundamental as measuring 
the sign of the Sivers asymmetry in the Drell-Yan process.

%
%
\section{Analytical results}
\begin{figure}[t]
\begin{center}
\includegraphics[width=7.5cm]{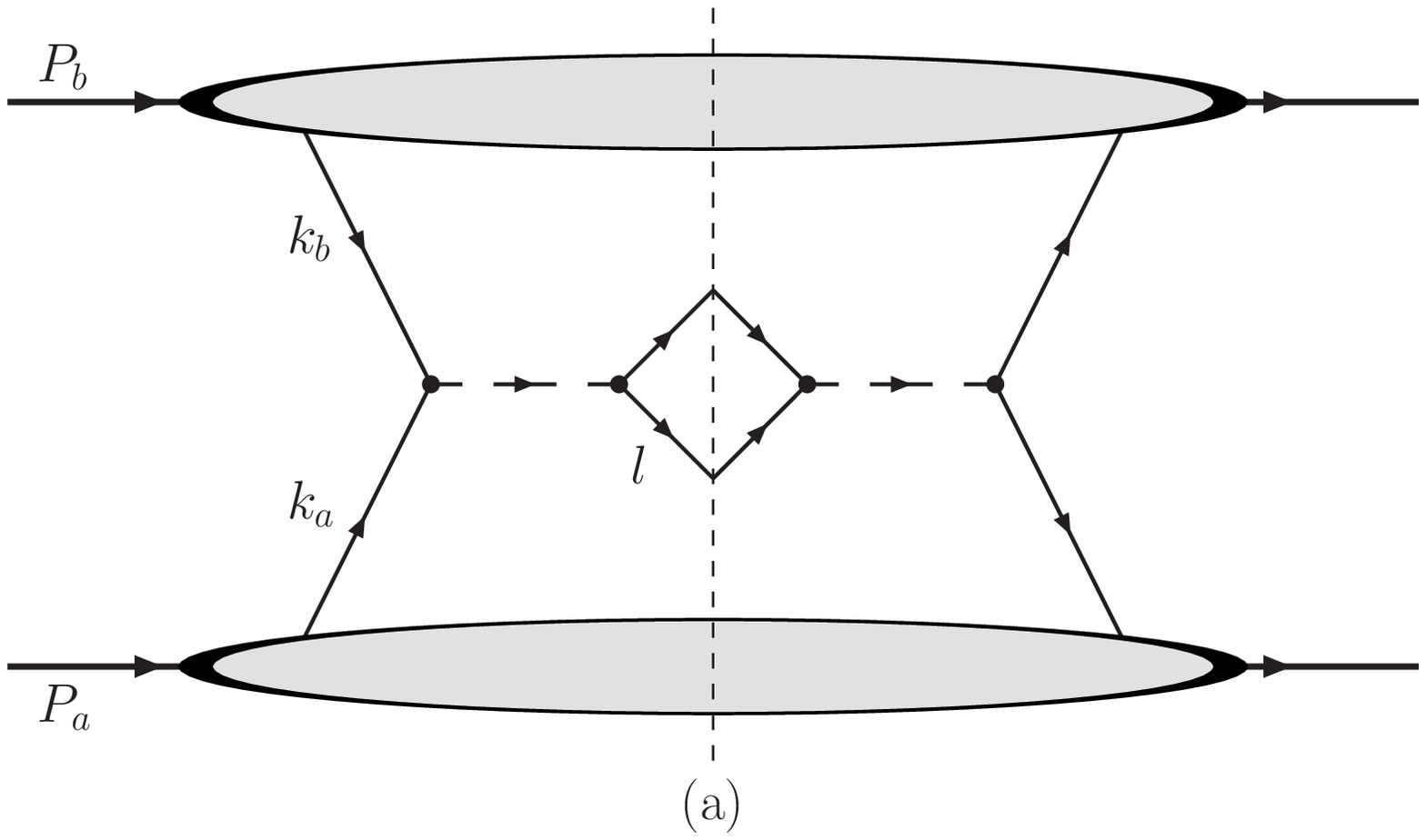}
\hskip 1.0cm
\includegraphics[width=7.5cm]{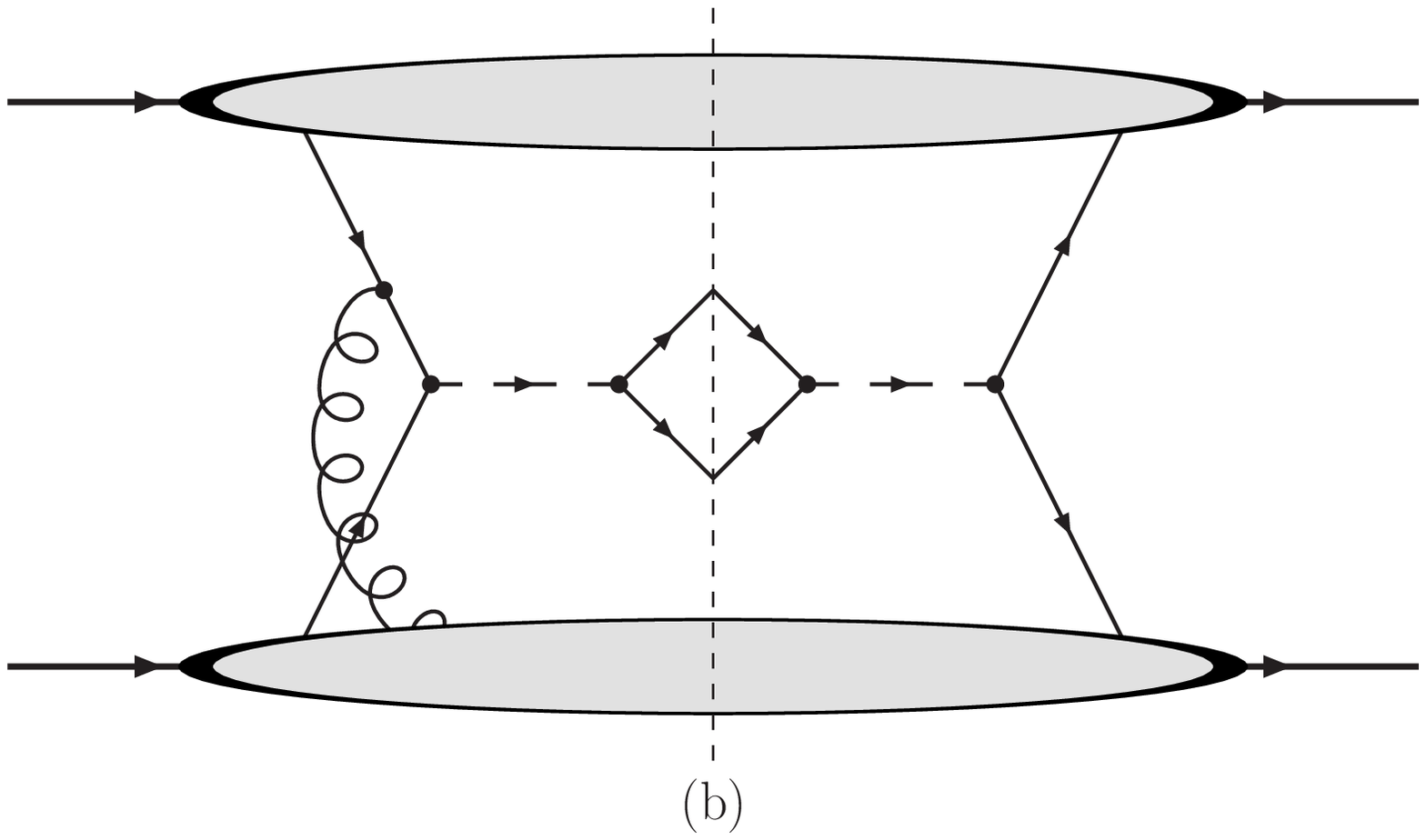}
\caption{Diagram (a): parton model representation for $p \, p \to l^{\pm} \, X$, where
the lepton is produced in the decay of a $W$-boson.
The final state (anti-)neutrino goes unobserved. 
Diagram (b): contribution from quark-gluon-quark correlation. This diagram, together
with its Hermitian conjugate which is not displayed, needs to be taken into account
when computing twist-3 observables.}
\label{f:diagram}
\end{center}
\end{figure}

\noindent
We start by fixing the kinematical variables for the process 
$p \, p \to l^{\pm} \, X$, and assign 4-momenta to the particles according to
\begin{equation}
p(P_a) + p(P_b) \to l^{\pm}(l) + X \,.
\end{equation}
By means of these momenta we specify a coordinate system through 
$\hat{e}_z = \hat{P_a} = - \hat{P_b}$, 
$\hat{e}_x = \hat{l}_{T}$ (with $\vec{l}_{T}$ representing the transverse momentum of 
the jet), and $\hat{e}_y = \hat{e}_z \times \hat{e}_x$.
Mandelstam variables are defined by
\begin{equation} \label{e:mandel_1}
s = (P_a + P_b)^2 \,, \qquad
t = (P_a - l)^2 \,, \qquad 
u = (P_b - l)^2 \,,
\end{equation}
while on the partonic level one has
\begin{equation} \label{e:mandel_2}
\hat{s} = (k_a + k_b)^2 = x_a x_b s\,, \qquad
\hat{t} = (k_a - l)^2 = x_a t\,, \qquad 
\hat{u} = (k_b - l)^2 = x_b u \,,
\end{equation}
where $k_a$ and $k_b$ denote the momentum of the active quark/antiquark in the 
protons; see also Fig.~\ref{f:diagram}(a).
The momentum fraction $x_a$ characterizes the (large) plus-momentum of the 
quark/antiquark in the proton moving along $\hat{e}_z$ through 
$k_a^+ = x_a P_a^+$.\footnote{For a generic 4-vector $v$, we define light-cone 
coordinates according to $v^{\pm} = (v^0 \pm v^3) / \sqrt{2}$ and 
$\vec{v}_T = (v^1,v^2)$.}
Likewise, one has $k_b^- = x_b P_b^-$.
The relation $\hat{s} + \hat{t} + \hat{u} = 0$ implies 
\begin{equation} \label{e:mom_rel}
x_a = - \frac{x_b u}{x_b s + t}
= \frac{x_b \sqrt{s} \, l_{T} \, e^{\eta}}
       {x_b s - \sqrt{s} \, l_{T} \, e^{-\eta}} \,.
\end{equation}
In the second step in~(\ref{e:mom_rel}) we express $x_a$, for a given $\sqrt{s}$, 
through $l_{T} = |\vec{l}_{T}|$ and the pseudo-rapidity $\eta = - \ln \tan(\vartheta/2)$ 
of the lepton, since transverse momenta and (pseudo-)rapidities are commonly used to 
describe the kinematics of a final state particle in proton-proton collisions.

Next, we turn to the polarization observables for $p \, p \to l^{\pm} \, X$, 
which we compute in the collinear factorization framework. 
As already mentioned, we focus on the situation when one proton is transversely
polarized, while the other is either unpolarized or longitudinally polarized.
One finds the following expression for the cross 
section:\footnote{Polarization degrees are suppressed in the cross section 
formula~(\ref{e:master}).}
\begin{eqnarray} \label{e:master}
\lefteqn{l^0 \, \frac{d^3 \sigma}{d^3 l} = 
\frac{\alpha_{em}^2}{12 \,s \sin^4 \vartheta_w} \, \sum_{a,b} \, |V_{ab}|^2  
\int_{x_b^{min}}^1 \frac{dx_b}{x_a x_b} \, \frac{1}{x_b s + t} \, 
\bigg\{ \, H^{ab} \, f_1^a(x_a) \, f_1^b(x_b)}
\nonumber \\
&& 
 + \, 2 \pi M \, \varepsilon_T^{ij} l_{T}^i S_{aT}^j \, \tilde{H}^{ab} \,
 \bigg[ \, 
 \bigg(T_{F}^a(x_a,x_a) - x_a \, \frac{d}{dx_a} \, T_{F}^a(x_a,x_a) \bigg) 
 + K(\hat{s}) \, T_F^a(x_a,x_a) \bigg] f_1^b(x_b) 
\nonumber \\
&& 
 + \, 2M \, \vec{l}_T \cdot \vec{S}_{aT} \, \tilde{H}^{ab} \, 
 \bigg[ \, 
 \bigg(\tilde{g}^a(x_a) - x_a \, \frac{d}{dx_a} \, \tilde{g}^a(x_a) \bigg) 
 + K(\hat{s}) \, \tilde{g}^a(x_a) 
 + 2 x_a \, g_T^a(x_a) \bigg] f_1^b(x_b) 
\nonumber \\
&& 
 - \, 2 \pi M \, \lambda_b \, \varepsilon_T^{ij} l_{T}^i S_{aT}^j  \, \tilde{H}^{ab} \, 
 \bigg[ \, 
 \bigg(T_{F}^a(x_a,x_a) - x_a \, \frac{d}{dx_a} \, T_{F}^a(x_a,x_a) \bigg) 
 + K(\hat{s}) \, T_F^a(x_a,x_a) \bigg] g_1^b(x_b) 
\nonumber \\
&& 
 - \, 2M \, \lambda_b \, \vec{l}_T \cdot \vec{S}_{aT} \, \tilde{H}^{ab} \, 
 \bigg[ \, 
 \bigg(\tilde{g}^a(x_a) - x_a \, \frac{d}{dx_a} \, \tilde{g}^a(x_a) \bigg) 
 + K(\hat{s}) \, \tilde{g}^a(x_a) 
 + 2 x_a \, g_T^a(x_a) \bigg] g_1^b(x_b)
\nonumber \\ 
&& 
 + \, \ldots \bigg\} \,,
\\
&& \textrm{with} \quad K(\hat{s}) =  \frac{2 M_W^2 (\hat{s} - M_W^2 - \Gamma_W^2)}
        {(\hat{s} - M_W^2)^2 + M_W^2 \Gamma_W^2} \,. 
\nonumber
\end{eqnarray}
In Eq.~(\ref{e:master}), $\vartheta_w$ is the weak mixing angle, $V_{ab}$ is a CKM
matrix element, $M$ is the proton mass, $M_W$ is the $W$-mass and $\Gamma_W$ its
decay width.
We also use $\varepsilon_T^{ij} \equiv \varepsilon^{-+ij}$ with 
$\varepsilon^{0123} = 1$.
The transverse spin vector of the proton moving along $\hat{e}_z$ is denoted by
$\vec{S}_{aT}$, whereas $\lambda_b$ represents the helicity of the second proton. 
The lower limit of the $x_b$-integration is given by $x_b^{min}=-t/(s+u)$.
One can project out the four spin-dependent components of the cross section 
in~(\ref{e:master}), in order, through
\begin{eqnarray} \label{e:sigma_tu_e}
\sigma_{TU}^e & = & \frac{1}{4} 
 \Big( \big[\sigma(\uparrow_{y},+) - \sigma(\downarrow_{y},+) \big] 
     + \big[\sigma(\uparrow_{y},-) - \sigma(\downarrow_{y},-) \big] \Big) \,,
\\ \label{e:sigma_tu_o}
\sigma_{TU}^o & = & \frac{1}{4}
 \Big( \big[\sigma(\uparrow_{x},+) - \sigma(\downarrow_{x},+) \big] 
     + \big[\sigma(\uparrow_{x},-) - \sigma(\downarrow_{x},-) \big] \Big) \,,
\\ \label{e:sigma_tl_o}
\sigma_{TL}^o & = & \frac{1}{4}
 \Big( \big[\sigma(\uparrow_{y},+) - \sigma(\downarrow_{y},+) \big] 
     - \big[\sigma(\uparrow_{y},-) - \sigma(\downarrow_{y},-) \big] \Big) \,,
\\ \label{e:sigma_tl_e}
\sigma_{TL}^e & = & \frac{1}{4} 
 \Big( \big[\sigma(\uparrow_{x},+) - \sigma(\downarrow_{x},+) \big] 
     - \big[\sigma(\uparrow_{x},-) - \sigma(\downarrow_{x},-) \big] \Big) \,.
\end{eqnarray}
In these formulas, '$\uparrow_{x/y}$' ('$\downarrow_{x/y}$') denotes transverse 
polarization along $\hat{e}_{x/y} \, (-\hat{e}_{x/y})$ for the proton moving 
in the $\hat{e}_z$-direction, whereas '$+$' and '$-$' represent the helicities of 
the second proton.

The dots in Eq.~(\ref{e:master}) indicate longitudinal single spin and double spin 
observables, as well as four possible correlations for double transverse 
polarization.
In collinear factorization, the latter are at least twist-4 effects in the
Standard Model.
Note that double transverse polarization observables for $W$-production were also 
discussed in connection with potential physics beyond the Standard Model 
(see~\cite{Rykov:1999ru,Boer:2010mc} and references therein).

We computed the (twist-2) unpolarized cross section in the first line of~(\ref{e:master})
on the basis of diagram (a) in Fig.~\ref{f:diagram} by applying the collinear 
approximation to the momenta $k_a$ and $k_b$ of the active partons.
The result contains the ordinary unpolarized quark distribution $f_1^a$ for a quark 
flavor $a$.
The hard scattering coefficients $H^{ab}$ and $\tilde{H}^{ab}$ in Eq.~(\ref{e:master}), 
expressed through the partonic Mandelstam variables in~(\ref{e:mandel_2}), read
\begin{equation} \label{e:hard}
H^{ab} = \frac{\hat{u}^2}{(\hat{s} - M_W^2)^2 + M_W^2 \Gamma_W^2} \,, \qquad
\tilde{H}^{ab} = \frac{1}{\hat{u}} \, H^{ab} \,, \quad
\textrm{for} \;\; ab = d\bar{u}, \; s\bar{u}, \; \bar{d}u, \; \bar{s}u \,. 
\end{equation}
In Eq.~(\ref{e:hard}), one has to replace $\hat{u}$ by $\hat{t}$ for
$ab = \bar{u}d, \; \bar{u}s, \; u\bar{d}, \; u\bar{s}$.

The four cross sections in~(\ref{e:sigma_tu_e})--(\ref{e:sigma_tl_e}) 
represent twist-3 observables.
Calculational details for such observables in collinear factorization can be found 
in various papers; see, e.g., 
Refs.~\cite{Qiu:1991pp,Eguchi:2006mc,Kouvaris:2006zy,Yuan:2008it,Kang:2008qh,Zhou:2009jm}.
We merely mention that one has to expand the hard scattering contributions around 
vanishing transverse parton momenta.
While for twist-2 effects only the leading term of that expansion matters, in the case
of twist-3 the second term is also relevant.
In addition, the contribution from quark-gluon-quark correlations, as displayed in
diagram (b) in Fig.~\ref{f:diagram}, needs to be taken into consideration.
The sum of all the terms can be written in a color gauge invariant form, which provides
a consistency check of the calculation.

The quark-gluon-quark correlator showing up in $\sigma_{TU}^e$ and $\sigma_{TL}^o$ 
is the aforementioned ETQS matrix element 
$T_F^a(x,x)$~\cite{Efremov:1981sh,Efremov:1984ip,Qiu:1991pp}.
The peculiar feature of this object is the vanishing gluon momentum --- that's why 
it is also called ``soft gluon pole matrix element''.
If the gluon momentum becomes soft one can hit the pole of a quark propagator in
the hard part of the process, providing an imaginary part (nontrivial phase) which, 
quite generally, can lead to single spin 
effects~\cite{Efremov:1981sh,Efremov:1984ip,Qiu:1991pp}.
Note also that in our lowest order calculation no so-called soft fermion pole 
contribution (see~\cite{Koike:2009ge} and references therein) emerges. 
For $\sigma_{TU}^o$ and $\sigma_{TL}^e$ another quark-gluon-quark matrix element --- 
denoted as $\tilde{g}^a$; see, in particular, 
Refs.~\cite{Eguchi:2006qz,Zhou:2008mz,Zhou:2009jm} --- appears, together with the 
familiar twist-3 quark-quark correlator $g_T^a$ (and, in the case of $\sigma_{TL}^o$,
together with the quark helicity distribution $g_1^a$).

We use the common definitions for $f_1$, $g_1$, and $g_T$. 
The quark-gluon-quark correlators $T_F$ and $\tilde{g}$ are specified according 
to\footnote{Note that in the literature different conventions for $T_F$ exist.} 
\begin{eqnarray} \label{e:defqgq_1}
- i \varepsilon_T^{ij} S_T^j \, T_F(x,x) & = & 
\frac{1}{2M} \int \frac{d\xi^- d\zeta^-}{(2\pi)^2} \, e^{i x P^+ \xi^-} \,
\langle P,S_T | \bar{\psi}(0) \, \gamma^+ \, 
              ig F^{+i}(\zeta^-) \, \psi(\xi^-) | P,S_T \rangle \,,
\\ \label{e:defqgq_2}
S_T^i \, \tilde{g}(x) & = & 
\frac{1}{2M} \int \frac{d\xi^-}{2\pi} \, e^{i x P^+ \xi^-} 
\nonumber \\
&& \hspace{1.0cm} \mbox{} \times
\langle P,S_T | \bar{\psi}(0) \, \gamma_5 \gamma^+ \,
              \bigg( i D_T^i - ig \int_0^\infty d\zeta^- F^{+i}(\zeta^-) \bigg) \,
              \psi(\xi^-) | P,S_T \rangle \,,
\end{eqnarray}
with $F^{\mu\nu}$ representing the gluon field strength tensor, and 
$D^{\mu} = \partial^{\mu} - i g A^{\mu}$ the covariant derivative.
Equations~(\ref{e:defqgq_1}) and~(\ref{e:defqgq_2}) hold in the light-cone gauge 
$A^+ = 0$, while in a general gauge Wilson lines need to be inserted between the 
field operators.

It is important that $T_F$ and $\tilde{g}$ are related to moments of TMDs.
To be explicit, one has~\cite{Boer:2003cm,Ma:2003ut,Zhou:2008mz,Zhou:2009jm}
\begin{eqnarray} \label{e:qgq_tmd_1}
\pi \, T_F(x,x) & = & 
- \int d^2k_T \, \frac{\vec{k}_T^2}{2M^2} \, 
              f_{1T}^{\perp}(x,\vec{k}_T^2)\Big|_{DIS} \,,
\\ \label{e:qgq_tmd_2}
\tilde{g}(x) & = & 
\int d^2k_T \, \frac{\vec{k}_T^2}{2M^2} \, g_{1T}(x,\vec{k}_T^2) \,,
\end{eqnarray}
where we use the conventions of 
Refs.~\cite{Mulders:1995dh,Boer:1997nt,Bacchetta:2006tn,Arnold:2008kf}
for the TMDs $f_{1T}^\perp$ and $g_{1T}$. 
In Eq.~(\ref{e:qgq_tmd_1}) we take into account that the Sivers function 
$f_{1T}^\perp$ depends on the process in which it is 
probed~\cite{Collins:2002kn,Brodsky:2002rv}.
In order to make numerical estimates we will exploit the relations 
in~(\ref{e:qgq_tmd_1}), (\ref{e:qgq_tmd_2}).

Finally, note that, due to the pure vector-axialvector coupling of the $W$-boson,
no chiral-odd parton correlator shows up in any of the four
spin correlations in~(\ref{e:master}), which makes those observables rather clean.
The situation is different if one considers single lepton production from the
decay of a virtual photon or of a $Z$-boson.

%
%
\section{Numerical results}
\noindent
Now we move on to discuss numerical results for the polarization observables
by limiting ourselves to the transverse single spin effects.
This means, we consider the two spin asymmetries $A_{TU}^e$ and $A_{TU}^o$, 
\begin{equation} \label{e:asymm}
A_{TU}^e = \frac{\sigma_{TU}^e}{\sigma_{UU}} \,, \qquad
A_{TU}^o = \frac{\sigma_{TU}^o}{\sigma_{UU}} \,,
\end{equation}
with $\sigma_{TU}^e$ and $\sigma_{TU}^o$ from Eq.~(\ref{e:sigma_tu_e}) 
and~(\ref{e:sigma_tu_o}), respectively, and $\sigma_{UU}$ denoting the unpolarized
cross section.
Note that the definition of $A_{TU}^e$ corresponds to the one of the transverse
SSA $A_N$, which has been extensively studied in one-particle inclusive production 
for hadron-hadron collisions; 
see~\cite{Adams:2003fx,Adler:2005in,Arsene:2008mi} for recent experimental
results from RHIC.

To compute $\sigma_{UU}$ we use the unpolarized parton densities from the
CTEQ6-parameterization~\cite{Pumplin:2002vw}.
For the ETQS matrix element we use the relation~(\ref{e:qgq_tmd_1}) between $T_F$ 
and the Sivers function, and take $f_{1T}^\perp$ from the recent fit provided in 
Ref.~\cite{Anselmino:2008sga} on the basis of data from semi-inclusive DIS.
(For experimental studies of the Sivers effect we refer 
to~\cite{Airapetian:2004tw,Alexakhin:2005iw}, while extractions of the Sivers function 
from data can be found 
in~\cite{Anselmino:2008sga,Efremov:2004tp,Anselmino:2005nn,Vogelsang:2005cs,Collins:2005ie,Arnold:2008ap}.)
In the case of $A_{TU}^o$ one needs input for $g_T$ and $\tilde{g}$.
For $g_T$ we resort to the frequently used Wandzura-Wilczek 
approximation~\cite{Wandzura:1977qf} (see~\cite{Accardi:2009au} for a recent study 
of the quality of this approximation) 
\begin{equation} \label{e:appr_1}
g_T(x) \approx \int_x^1 \frac{dy}{y} \, g_1(y) \,,
\end{equation}
whereas for $\tilde{g}$ we use~(\ref{e:qgq_tmd_2}) and a Wandzura-Wilczek-type
approximation for the particular $k_T$-moment of $g_{1T}$ 
in~(\ref{e:qgq_tmd_2})~\cite{Metz:2008ib}, leading to
\begin{equation} \label{e:appr_2}
\tilde{g}(x) \approx x \int_x^1 \frac{dy}{y} \, g_1(y) \,.
\end{equation}
We mention that~(\ref{e:appr_2}) and a corresponding relation between chiral-odd 
parton distributions were used in~\cite{Kotzinian:2006dw,Avakian:2007mv} in order 
to estimate certain spin asymmetries in semi-inclusive DIS.
The comparison to data discussed in~\cite{Avakian:2007mv} looks promising, though 
more experimental information is needed for a thorough test of approximate 
relations like the one in~(\ref{e:appr_2}).
Measuring the SSA $A_{TU}^o$ could provide such a test.
The helicity distributions $g_1^a$ in~(\ref{e:appr_1}) and~(\ref{e:appr_2}) are 
taken from the DSSV-parameterization~\cite{deFlorian:2008mr}.
The transverse momentum of the lepton $l_T$ serves as the scale for the parton 
distributions.

The numerical estimates are for typical RHIC kinematics, i.e., 
$\sqrt{s} = 500 \, \textrm{GeV}$.
We present the asymmetries either as function of $\eta$ for fixed $l_T$ or 
vice versa.
\begin{figure}[t]
\begin{center}
\includegraphics[width=7.5cm]{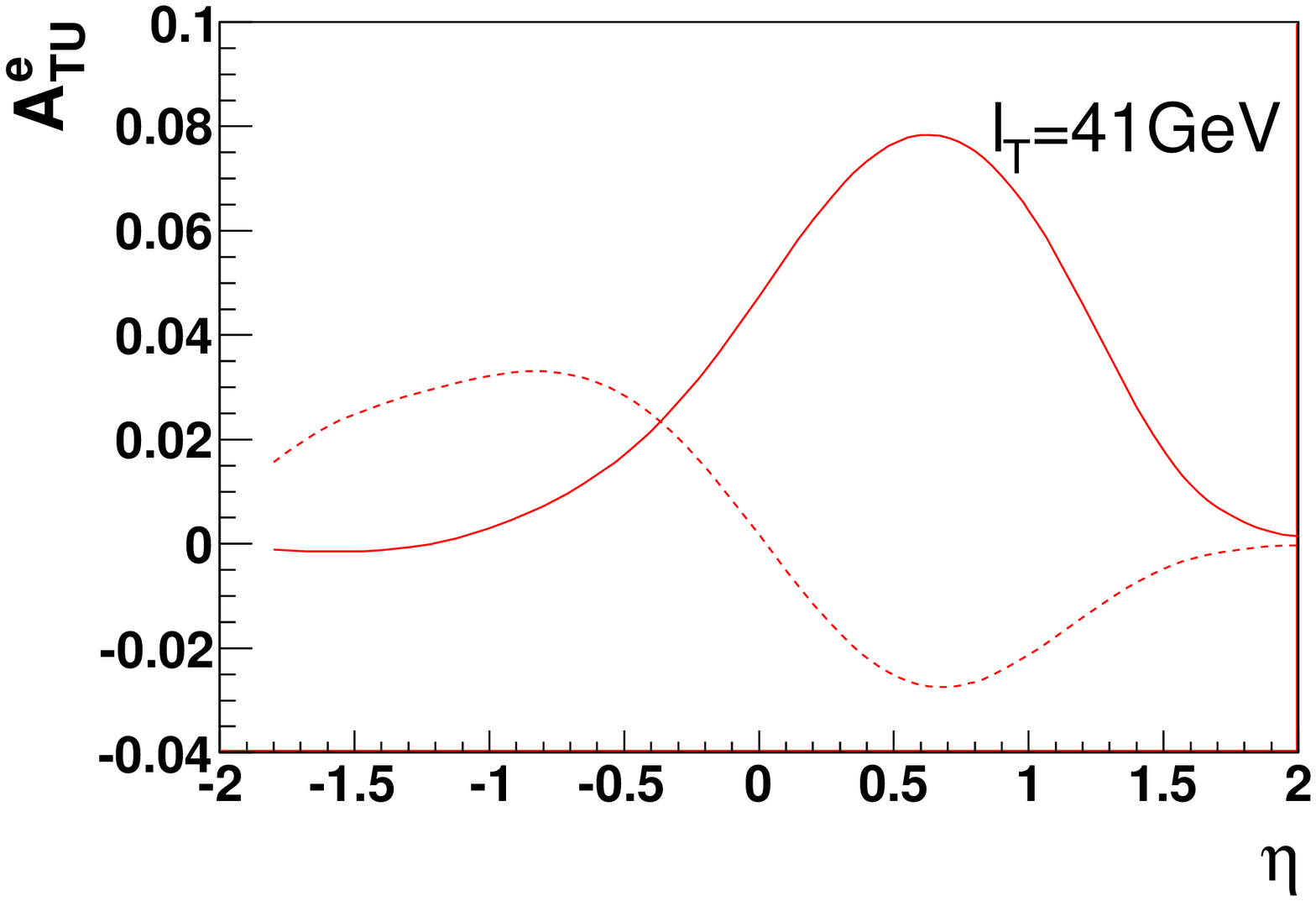}
\hskip 1.0cm
\includegraphics[width=7.5cm]{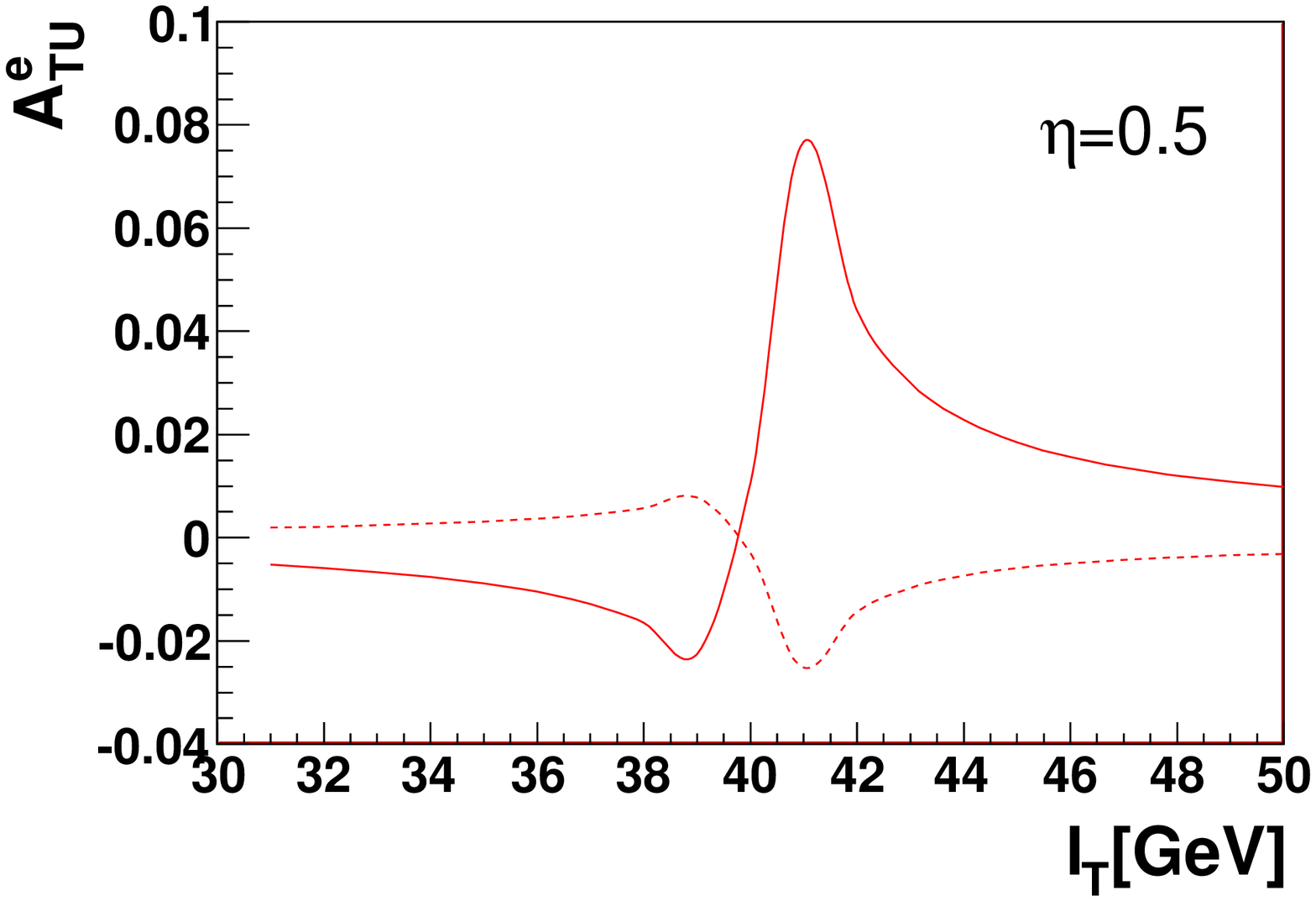}
\caption{$A_{TU}^e$ for $p \, p \to l^{\pm} \, X$ as a function of $\eta$ (left) and 
$l_T$ (right) for $\sqrt{s} = 500 \, \textrm{GeV}$.
The solid line is for $l^-$-production, and the dashed line is for $l^+$-production.}  
\label{f:atu_e}
\end{center}
\end{figure}

We start by discussing the parity-even asymmetry $A_{TU}^e$.
As shown in the right plot in Fig.~\ref{f:atu_e}, this observable is peaked around
$l_T \approx M_W/2$ --- a feature that does not depend on the value of $\eta$.
To be more precise, the peak is at $l_T = 41 \, \textrm{GeV}$, i.e., slightly above
$M_W/2$. 
The peak in the polarized cross section $\sigma_{TU}^e$ gets enhanced in the
asymmetry, because the unpolarized cross section drops rather fast when going
beyond $l_T = M_W/2$.
(As a side-remark we point out that the asymmetry in the peak region is 
completely dominated by the third term in the 2nd line in~(\ref{e:master}) 
containing the factor $K(\hat{s})$.)
Nevertheless, in this kinematical region we expect $A_{TU}^e$ to be measurable.
As discussed in the introduction, in this context it is important to recall that 
information on the sign of the asymmetry is already sufficient for a crucial test 
of our current understanding of transverse SSAs.

In particular in the peak region, the asymmetry is larger for $l^-$-production 
($W^-$-production) than for $l^+$-production, which is partly due to the rather 
large Sivers function for $d$-quarks obtained in the fit of 
Ref.~\cite{Anselmino:2008sga}.
The $l^-$-asymmetry and $l^+$-asymmetry come with opposite sign because the Sivers 
function for $u$-quarks and $d$-quarks have an opposite sign.
Note also that both asymmetries change sign as function of $l_T$.
Therefore, whether the sign of the asymmetry can be measured unambiguously may
critically depend on the $l_T$-resolution in the experiment.

As the $\eta$-dependence of $A_{TU}^e$ in left plot in Fig.~\ref{f:atu_e} shows, the 
asymmetry is maximal in the positive $\eta$ range, when a large-$x$ parton from the 
polarized proton participates in the hard scattering.
Obviously, by integrating over a suitable $\eta$-range one may optimize between 
magnitude of the asymmetry on the one hand and the size of the statistical error bars 
on the other.
Moreover, it is worthwhile to mention that the contributions from the antiquark 
Sivers functions are not negligible in the backward region. 
(Here we refer to a corresponding discussion on the Sivers asymmetry in the Drell-Yan 
process for proton-proton collisions in~\cite{Collins:2005rq}, where the strong 
sensitivity to the Sivers function for antiquarks was already pointed out.)
\begin{figure}[t]
\begin{center}
\includegraphics[width=7.5cm]{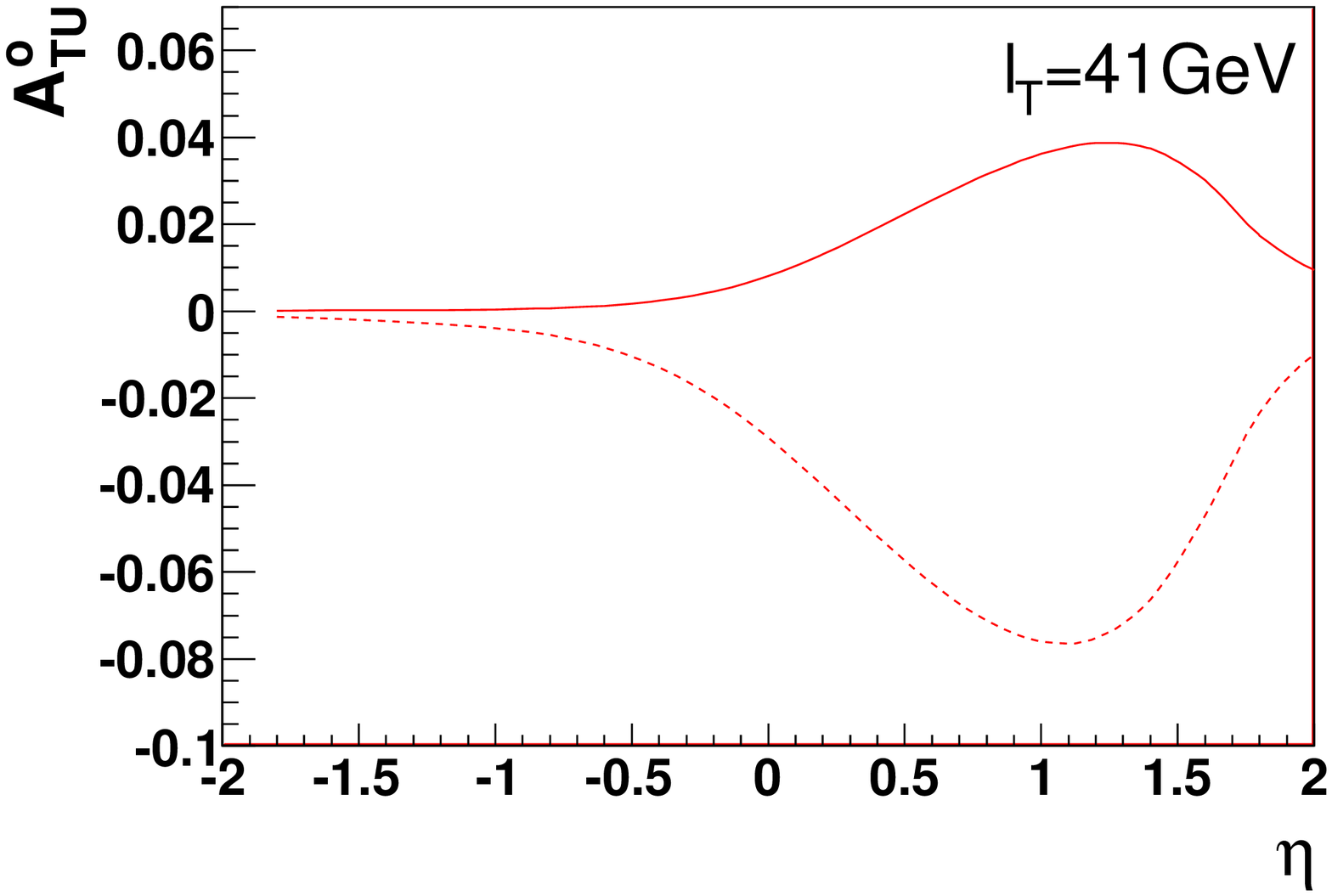}
\hskip 1.0cm
\includegraphics[width=7.5cm]{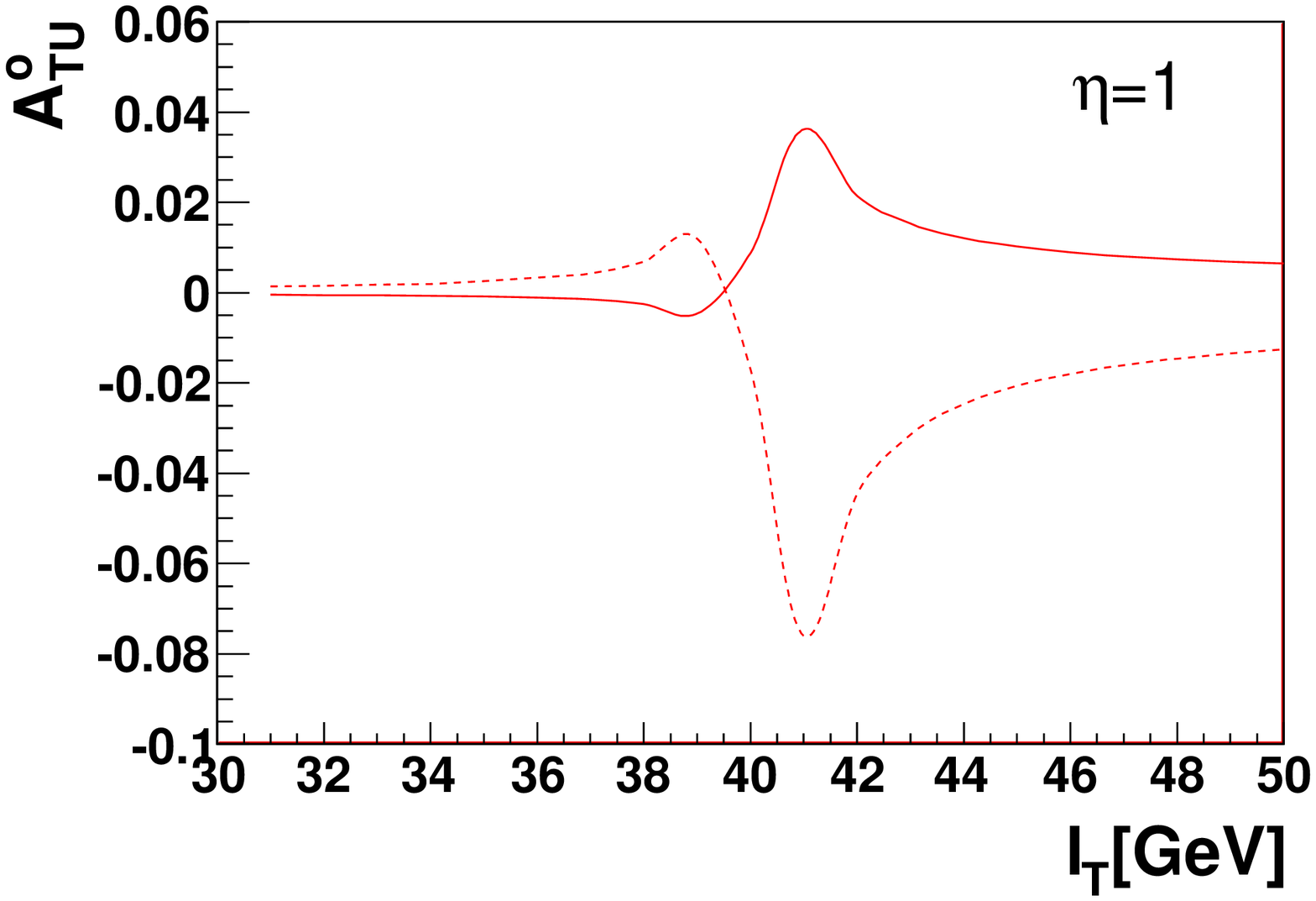}
\caption{$A_{TU}^o$ for $p \, p \to l^{\pm} \, X$ as a function of $\eta$ (left) and 
$l_T$ (right) for $\sqrt{s} = 500 \, \textrm{GeV}$.
The solid line is for $l^-$-production, and the dashed line is for $l^+$-production.}  
\label{f:atu_o}
\end{center}
\end{figure}

It is also interesting that for both $l^+$-production and $l^-$-production 
the overall magnitude of $A_{TU}^e$ is very similar to the predictions presented in 
Ref.~\cite{Kang:2009bp}, where TMD-factorization was used. 

Let us now turn to the parity-odd transverse SSA $A_{TU}^o$, which is displayed in 
Fig.~\ref{f:atu_o}. 
Again, this asymmetry has a pronounced peak at $l_T = 41 \, \textrm{GeV}$, and it is
largest for $l^+$-production (up to about $8 \, \%$).
As outlined above, our prediction for $A_{TU}^o$ is based on the Wandzura-Wilczek-type
approximation leading to~(\ref{e:appr_2}), which probably represents the most uncertain
part of our calculation.
Nevertheless, the asymmetry should be within experimental reach.
Like in the case of the parity-even SSA, also $A_{TU}^o$ is almost entirely determined 
by the $K(\hat{s})$-term in the 3rd line in~(\ref{e:master}). 
This implies that, due to the relation~(\ref{e:qgq_tmd_2}), it gives rather clean 
access to the TMD $g_{1T}$, which so far is experimentally unconstrained.
Therefore, in any case, a measurement of $A_{TU}^o$ would provide very interesting 
new information.

%
%
\section{Summary}
\noindent
We have studied transverse spin asymmetries for the process $p \, p \to l^{\pm} \, X$, 
where the lepton is produced in the decay of a $W$-boson.
If one of the protons is transversely polarized, and the other is either unpolarized
or longitudinally polarized, there exist two parity-even and two parity-odd
spin asymmetries.
We computed these asymmetries in collinear twist-3 factorization at the level 
of Born diagrams.
Moreover, for the two transverse single spin asymmetries $A_{TU}^e$ and $A_{TU}^o$
--- defined through Eq.~(\ref{e:asymm}) and~(\ref{e:sigma_tu_e}),~(\ref{e:sigma_tu_o}) 
--- we made numerical estimates for typical kinematics at 
RHIC ($\sqrt{s} = 500 \, \textrm{GeV}$). 
In the following we summarize our main results:
\begin{itemize}
\item The analytical results for all four spin-dependent cross sections are given
by two particular quark-gluon-quark correlators, which have a direct relation
to transverse momentum dependent parton distributions: the Sivers function $f_{1T}^\perp$ 
and the TMD $g_{1T}$; see Eqs.~(\ref{e:qgq_tmd_1}),~(\ref{e:qgq_tmd_2}).
Measuring these observables could therefore provide new information on the structure 
of the proton that goes beyond the collinear parton model.
\item The parity-even SSA $A_{TU}^e$ is largest for $l^-$-production (up to
about $8 \, \%$), and it is peaked for transverse momenta $l_T$ of the lepton 
slightly above $M_W/2$.
(Actually, all the asymmetries studied in this Letter are significant only in a 
relatively narrow region around $l_T \approx M_W/2$.)
Measuring the sign of this asymmetry can, in essence, provide an as crucial
test as measuring the sign of the Sivers asymmetry in Drell-Yan would do: 
it can test our present understanding of the underlying dynamics of transverse 
SSAs and at the same time check an important ingredient of TMD-factorization,
namely the influence of the Wilson-line which is generated by the interaction
between the active partons and the remnants of the protons.
(For related work we refer 
to~\cite{Collins:2002kn,Brodsky:2002rv,Efremov:2004tp,Bacchetta:2007sz,Kang:2009bp,Anselmino:2009st,Kang:2009sm}.)
\item To the best of our knowledge the parity-odd SSA $A_{TU}^o$ was 
never before explored in the literature.
We find $A_{TU}^o$ to be largest for $l^+$-production (also up to about $8 \, \%$,
like $A_{TU}^e$ for $l^-$-production).
This observable is directly related to (a moment of) the TMD $g_{1T}$, for which
at this time no experimental information exists. 
\end{itemize}
In general, we believe that $W$-physics for polarized proton-proton collisions is 
very promising not only in the case of longitudinally polarized protons, but has 
also a considerable discovery potential for transverse polarization.
\\[0.5cm]
%
%
\noindent
{\bf Acknowledgments:} 
We thank Zhongbo Kang, Jianwei Qiu, and Feng Yuan for helpful discussions.
A.M. acknowledges the support of the NSF under Grant No.~PHY-0855501.

%
%

\end{document}